# Deployable Networks for Public Safety in 5G and Beyond: A Coverage and Interference Study

Zhiqiang Qi, Adrián Lahuerta-Lavieja, Jingya Li, Keerthi Kumar Nagalapur

Ericsson

Contact: {zhiqiang.qi, adrian.lahuerta.lavieja, jingya.li, keerthi.kumar.nagalapur}@ericsson.com

*Abstract*— Deployable networks are foreseen to be one of the key technologies for public safety in fifth generation (5G) mobile communications and beyond. They can be used to complement the existing public cellular networks to provide temporary and on-demand connectivity in emergency situations. However, operating deployable networks in coexistence with public cellular networks can be challenging from an interference perspective. To gain insights on the deployment strategy for deployable networks, in this article, we present an extensive numerical study of coverage and interference analysis, considering four different co-existence scenarios and different types of deployable base stations (BSs), i.e., BS on a truck and BS on an Unmanned Aerial Vehicle (UAV). Our simulation results show that deploying deployable BSs in rural scenarios can provide good coverage to meet the service requirement for mission critical (MC) users. In addition, the interference impact is only substantial when the deployable and public networks are close to each other. Finally, allowing the MC users to access the public network can be of vital importance to guarantee their service when the interference level between public and deployable network is very high.

## I. INTRODUCTION

Fifth generation (5G) and beyond mobile communications are going to transform our lives and make them safer. For public safety, where the availability and reliability of communication can translate into life or death, New Radio (NR), the 5G's wireless access technology, has been developed and enhanced to support mission critical (MC) communications [1]. These features are, for example, repetition and relaying for extended coverage, multicast and broadcast for efficient group communications, access control and random access for prioritizing MC traffic, and enhanced positioning for first responders [2].

Despite these powerful features, coverage and capacity provided by the existing mobile networks at a particular geographical disaster-struck area may not be sufficient to meet the stringent requirements of public safety operations. Consequently, to ensure real-time and reliable connectivity when and where it counts, there has been a growing interest in public safety communities to use deployable networks for providing temporary on-demand coverage, capacity, and redundancy during MC situations. See, for example, [3], [4].

Deployable base stations (BSs) are portable BS. Commonly, they are installed on vehicles such that they can be brought as close as possible to the target geographical area, i.e., where first responders will be carrying out search and rescue missions. The most common vehicles are land-based vehicles, such as trucks, also known as cells-on-wheels; and unmanned aerial vehicles (UAVs), such as balloons and drones, also known as cell-on-wings. In literature we can also find examples of deployable BSs installed on other vehicles, e.g., on crane cars [5] or on towable person lifts [6]. The commonality between the two last examples is the higher heights at which the BS is deployed for providing coverage to a larger geographical area. UAVs, with their ability to fly at higher heights, provide further flexibility in terms of deployment. Note that, in the rest of the paper we will often use the terms truck and UAV to refer to cell-on-wheels and cell-on-wings, respectively, for simplicity.

Deployable BSs are not new to NR. They were present with Long Term Evolution (LTE) and typically relied on satellite backhaul, which is not the most desirable option due to cost- and performance-related reasons [3], [4]. Integrated access and backhaul (IAB), a new feature introduced in NR, offers an innovative type of wireless backhaul to not only densify networks but also to configure a flexible network topology. The potential of IAB goes beyond public safety due to its flexibility and scalability with use cases in, for example, urban areas with high demands on capacity and speed, as well as in suburban areas with fiber-like fixed wireless access (FWA) services [7].

Exploiting the opportunities provided by current and future cellular systems is essential for enabling a high-performance public safety infrastructure [8]. For example, the use of UAVs together with conventional cellular networks for 5G NR will not only help meet the stringent requirements of MC traffic but will also empower an increase in the energy efficiency of the network [9]. Another example, using several 5G and beyond features, such as IAB and millimeter-wave (mm-wave) spectrum, points out to the substantial gains of dynamic positioning of deployable BSs compared to rigid deployments [10]. In this context, we have seen a significant presence of studies investigating optimal UAV positioning/deployment by means of machine learning algorithms, e.g., [11]–[15]. Recent works have also looked into, e.g., latency [16], spectrum sharing [17]–[19], network slicing [19], [20], security [21], and energy efficiency [22] aspects.

In this paper, we present an extensive system level simulation study to investigate the coverage and interference performance when deploying a deployable network in coexistence with a public cellular network. Four different deployable and public network coexistence scenarios are considered, where a public-network-only scenario is used as the baseline to assess the benefits and drawbacks of using a deployable network in serving first responders during a MC situation.

The rest of the paper is structured as follows. Section II introduces the system model considered in the study. Section III presents coverage and interference results of the public and deployable networks. Finally, Section IV summarizes our findings and concludes this paper.

## II. SYSTEM MODEL

We consider a multi-network coexistence scenario, which consists of a public network and a deployable network, as shown in Fig.1. A group of MC users are uniformly distributed in an emergency area, i.e., the central area of the deployment map shown in Fig. 1, which is located approximately 10 km away from the two nearest macro BSs of the public network. Due to very large distance between the MC users and the nearest two macro BSs, these MC users have very limited or no connectivity to the public network. Thereby, a deployable network is setup in the MC area to provide temporary on-demand coverage to the MC users.

Two types of deployable BSs are considered in this paper: cell-on-wheels (truck) and cell-on-wings (UAV). Taking truck-mounted BS as an example, cell-on-wheels typically has less constraints in cruising duration and they can transmit with higher power to provide a relatively large coverage area. However, the placement of cell-on-wheels can be less flexible for MC operations in rural areas with complicated environments, such as forest firefighting, and search and rescue in mountainous regions. On the other hand, cell-on-wings, e.g., a UAV carrying a BS, can be deployed in a more flexible way with a higher mobility. The drawback of UAV-based deployable BSs is that they typically have stricter constraints on hardware size, weight, and power consumption, leading to low transmit power and low operation time. Taking these aspects into account, in our simulation setup, cell-on-wings are configured with only a single sector per BS while the cell-on-wheels are configured with three sectors per BS.

In this study, we assumed that the deployable network is operating in a standalone mode, meaning that there is no communication between deployable and public networks. Users are divided into two groups, i.e., normal users and MC users. Normal users are located within coverage area of the two macro BSs, and they are not allowed to connect to the deployable network, they can only be served by macro BSs. MC users connect to both deployable and public networks and MC select either a macro BS or a deployable BS as its serving node based on a certain cell selection criterion, e.g., the radio link quality.

In a MC scenario, it is important that all the MC users in the MC area are provided with the guaranteed bitrate to support their required MC service. Therefore, rather than pursuing high throughput for only a subset of the MC users, we focus on metrics like requested traffic and served traffic for all MC users. Here, the pre-configured requested traffic is defined as the traffic requirement $T_{Req}$ for certain services (MC or non-MC type). If the link quality between a user and its serving BS is not good enough to provide a minimum requested data traffic, then, this user will be dropped, leading to dropped traffic, which is represented by $T_{Dropped}$. In some cases, even with good link quality, some traffic may still not be delivered due to insufficient radio resources, and this part of traffic is modelled as blocked, denoted by $T_{Blocked}$. Then, the actual delivered traffic $T_{Served}$ can be calculated as:

$$T_{Served} = T_{Req} - T_{Dropped} - T_{Blocked}. \quad (1.1)$$

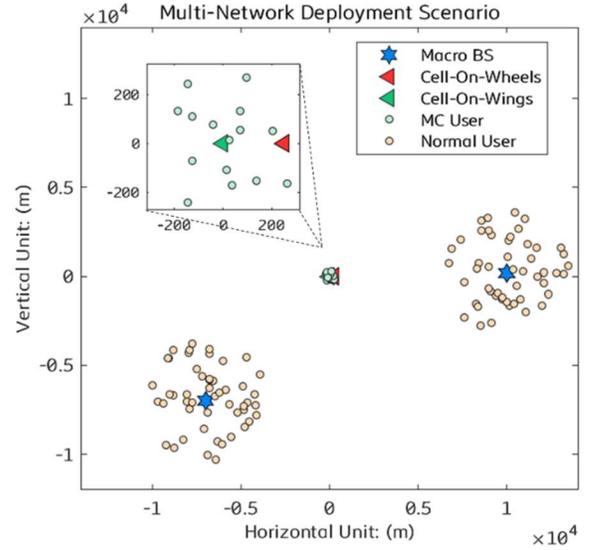

Fig. 2. Geometry of a deployable network (center of the figure and magnified) in coexistence with a public cellular network.

Note that, a user will not be served with more traffic than what is required, even if its Signal to Noise Ratio (SNR) or Signal to Interference plus Noise Ratio (SINR) is higher than what is needed for transmitting the required $T_{Req}$ bits.

## III. NUMERICAL RESULTS

In this section, the numerical results for coverage and interference analysis are introduced. Based on the system model discussed in Section II, the simulation is performed using a MATLAB-based system simulator using the simulation parameters shown in Table 1. A rural-macro propagation model [23][1] is used as the propagation model. The cell gain consists of antenna gain and path gain, where the path gain includes both pathloss and shadowing gain.

To investigate the potential of using deployable BSs to provide different MC services with different requirements, two traffic patterns are considered, namely, downlink-heavy and uplink-heavy, as shown in Table 1. Due to space constraints, figures showing obvious and similar results are omitted and we only state the conclusion.

### A. Coverage Analysis

To better reflect deployable BSs' capability to improve coverage performance, four scenarios are evaluated in this subsection:
1. Macro-Only scenario: In this scenario, deployable BSs are not considered, and all users can only connect to macro BSs. We can observe the performance of MC users without introducing deployable BSs.

Table 1 Simulation Parameters

| Parameter | Value | | |
|---|---|---|---|
| | Macro BS | Cell-on-Wheels (Truck) | Cell-on-Wings (UAV) |
| Carrier Frequency | 700 MHz | | |
| Carrier Bandwidth | 10 MHz | | |
| Duplex | FDD | | |
| BS Antenna | 2Tx/2Rx | | |
| BS Transmit Power | 49 dBm | 46 dBm (33 dBm for low deployable power case) | 40 dBm (24 dBm for low deployable power case) |
| BS Height | 32 m | 20 m | 25 m |
| Number of normal users | 50 around each Macro BS | | |
| Number of MC users | 15 | | |
| User Antenna | 2Tx/2Rx | | |
| User Transmit Power | 23 dBm | | |
| Macro Traffic Requirement | DL 1Mbps; UL 0.5Mbps | | |
| MC Traffic Requirement | Downlink heavy: DL 2 Mbps; UL 0.5 Mbps  Uplink heavy: DL 2 Mbps; UL 2 Mbps | | |

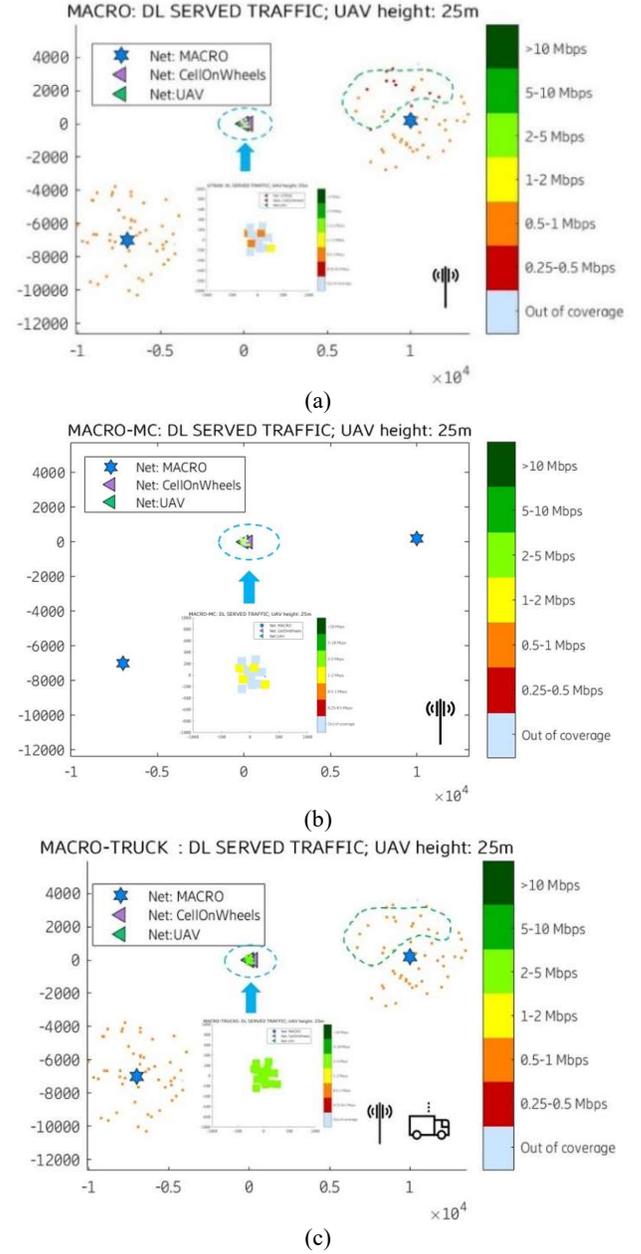

Fig. 2. DL served traffic in (a) Macro-Only, (b) Macro-MC, and (c) Macro-Truck scenarios.

2. Macro-MC scenario: This can be considered as a scenario where public network applies access and admission control to prioritize the MC traffic by blocking and pre-empting non-MC traffic. Through this scenario, we investigate whether more MC users can connect to the public network and be served the requested traffic, if all normal users are blocked from accessing the macro BSs.
3. Macro-Truck scenario: In this scenario, one cell-on-wheels is activated at the edge of the MC area to simulate the scenario of providing MC service in a remote forest.
4. Macro-UAV scenario: In this scenario, one cell-on-wings is activated above the center of the MC area.

By evaluating the performance in Macro-Truck and Macro-UAV scenarios, we can compare the difference between deploying cell-on-wheels and cell-on-wings. Specifically, when MC users are distributed in a forest and request MC services, both cell-on-wheels and cell-on-wings can be deployed to provide connectivity. However, a UAV can fly to the center of the forest while the truck can only drive to the edge.

Figs. 2(a), 2(b), and 2(c) show DL served traffic performance for Macro-Only, Macro-MC and Macro-Truck scenarios, respectively. The dot color in the figures denotes the exact value of served traffic as indicated in the color bar. For the Macro-Only scenario in Fig. 2(a), it can be observed that most of the MC users cannot be served the requested 2 Mbps traffic, as shown in the magnified subfigure. If admission control is applied for macro BS to block all normal users from accessing, the results shown in Fig. 2(b) show that most of the MC users still cannot be served the requested traffic in this Macro-MC scenario. Therefore, the limiting factor in using existing cellular network to serve MC UEs in this scenario is the poor radio link between MC users and macro cells, and macro-cell admission control cannot help. This shows the necessity of introducing deployable BSs to complement the existing public network to provide coverage to the MC users. If a cell-on-wheels with transmit power of 46 dBm is activated as shown in Fig. 2(c), then, all MC users can be served the requested traffic.

Additionally, by comparing the performance of normal users in these two scenarios, served traffic of some normal users, who are served by the macro BS on the right, improves after activating a deployable BS, as shown in the dashed-circled area in Figs. 2(a) and 2(c). This means that a deployable BS can help offload some MC traffic from the macro BS, which improves the performance of normal users by freeing up some macro BS resources. Based on the current configuration, there is not much difference in serving traffic with either cell-on-wheels or cell-on-wings. Hence, the results for Macro-UAV scenario will not be discussed here to avoid duplication.

To further evaluate the SINR performance, Figs. 3(a) and 3(b) show the DL SINR of all UEs for the considered Macro-Only and Macro-Truck scenarios, respectively. The x-axis of each subplot in Fig. 3 denotes the user indexes where
the first fifteen indexes represent the fifteen MC users and the rest denote normal users. It can be seen from Fig. 3(a) that for the Macro-Only scenario, the SINR values of MC users are quite low, mainly due to the bad channel quality between these MC users and their nearest Marco BS. When a cell-on-wheels is deployed close to the MC area, as shown in Fig. 3(b), the SINR values of MC users are significantly improved.

The high SNR values of MC UEs achieved by adding a cell-on-wheels as shown in Fig. 3(b) indicates the possibility of configuring a lower transmit power level from a deployable BS that can still fulfill the service requirement of MC UEs.

Fig. 4 illustrates the uplink performance of the considered scenarios. As in the downlink case, it is expected that introducing cell-on-wheels or cell-on-wings can also help to achieve the uplink traffic requirements for MC users. Despite of not being shown, our simulation results indicate that both options, i.e., cell-on-wheels and cell-on-wings, can meet the UL service requirement for all MC users if the required UL traffic is relatively low, e.g., 0.5Mbps. To further differentiate the capabilities of cell-on-wheels and cell-on-wings, an uplink-heavy traffic pattern is investigated, where the required uplink traffic performance in terms of served traffic in the Macro-Truck and Macro-UAV scenarios is indicated in Table I. As shown in Fig. 4(a), all MC users can be served 2 Mbps UL traffic if the deployable BS is cell-on-wheels. When the deployable BS is replaced by cell-on-wings, which only has a single sector, the 2 Mbps UL service requirement cannot be fulfilled, see Fig. 4(b). Therefore, the type of deployable BSs should be selected according to the targeted service requirement.

### B. Interference Analysis

In this subsection, we focus on the impact that interference between public and deployable network has on the traffic served to MC users. For this, we consider the geometry depicted in Fig. 1 together with scenario 4, i.e., the Macro-UAV scenario. In order to observe the impact of network interference, we vary the distance between one of the macro BSs (in this case, the macro BS in the left bottom corner in Fig. 1 at coordinates $(x, y) = (-7.5, -7.5)$ km) and the deployable BS by moving the MC area, the MC users, and the deployable BS such that the they are relatively static, see Fig. 5. Lastly, the movement of the MC area is done in the interval [0, 10]-km with a resolution of 10 meters.

Figs. 6(a), 6(b), 6(c), and 6(d) show served DL traffic per MC user in Mbps. Note that, the x-axis of the figures show the distance between the deployable BS and the macro base station. Hence, a distance of 10 km is equivalent to the scenario geometry shown in Fig. 1. In the four plots, for medium to large distances between public and private networks, i.e., from 5 km and farther, all (or almost all) MC users are served the requested (maximum) traffic, i.e., 2 Mbps. However, as we move the deployable BS closer to the macro BS, the effects of interference start to be more noticeable, especially when the deployable UAV uses the lower Tx power (24 dBm).

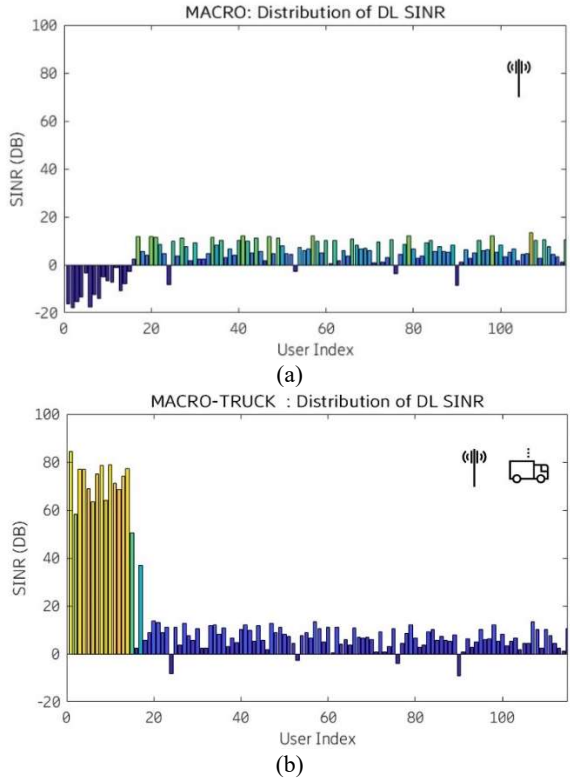

Fig. 3. DL SINR distribution in (a) Macro-Only and (b) Macro-Truck scenarios.

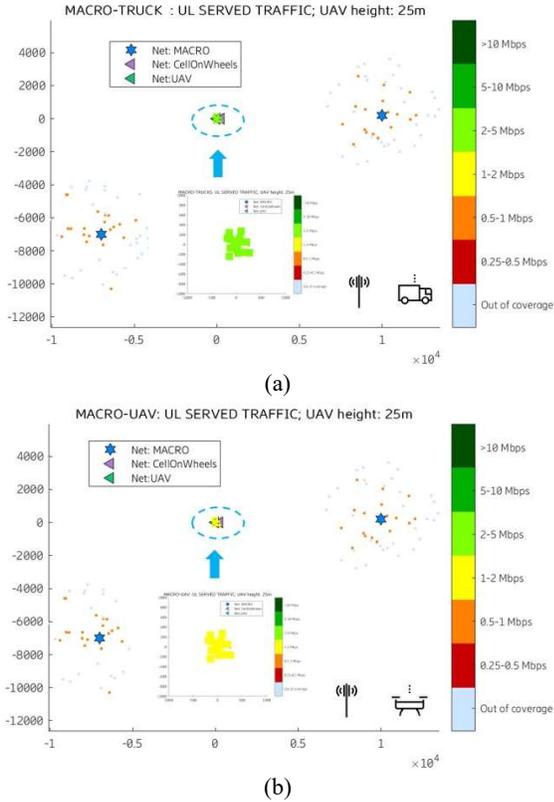

Fig. 4. Served traffic in the UL-Heavy case in (a) Macro-Truck and (b) Macro-UAV scenarios.

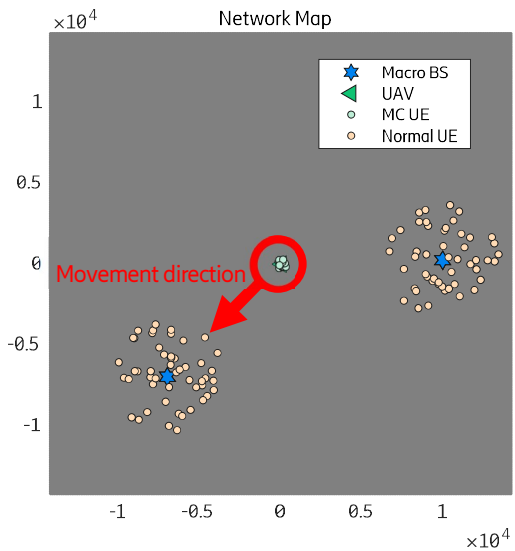

Fig. 5. Schematic drawing showing the movement direction of the MC area.

On the one hand, if the deployable BS is placed close to a macro BS (e.g., with a distance shorter than 1km) and the MC users can only access the deployable network, this has a major impact: almost no MC user is able to receive any traffic for distances shorter than 1 km due to the low SINR level. The low SINR resulting from high interference is due to the significant difference in the macro and deployable BSs Tx powers. This can be observed in Figs. 6(a) and 6(b).

On the other hand, when the MC users are allowed to receive service from the public network (if the quality of this link is better than the link quality to the deployable network), the MC users traffic is also degraded for shorter distances between macro and deployable BSs but to a lesser extent, receiving around 1Mbps in the worst cases. This can be observed in Figs. 6(c) and 6(d).

Furthermore, the results for the higher deployable BS Tx power, i.e., Figs. 6(a) and 6(c), and lower deployable BS Tx power, i.e., Figs. 6(b) and 6(d), show that MC users are less impacted by interference when the deployable BS uses a higher Tx power. The reason for this is as follows: for the same interference level, i.e., the signal transmitted by the macro BS to other users, the level of the signal of interest is higher due to the higher Tx power of the deployable BS. As a result, the MC users will connect to the public network for shorter distances to the macro BS as compared to the low power case.

Fig. 6 shows results for MC users only due to space limitations. However, normal UEs are also impacted due to interference between public and deployable BSs, i.e., when the deployable Bs is placed closer to normal users, they will be served less traffic due to interference. Moreover, the higher the deployable Tx power is, the more interference, and, hence, the worse QoS for normal users. Essentially, the deployable BS Tx power becomes a trade-off parameter that will impact the QoS of both MC and normal users. Hence, power-control mechanisms for interference-handling will need to be considered to find the optimal deployable BS Tx power in different MC scenarios, e.g., rural or urban.

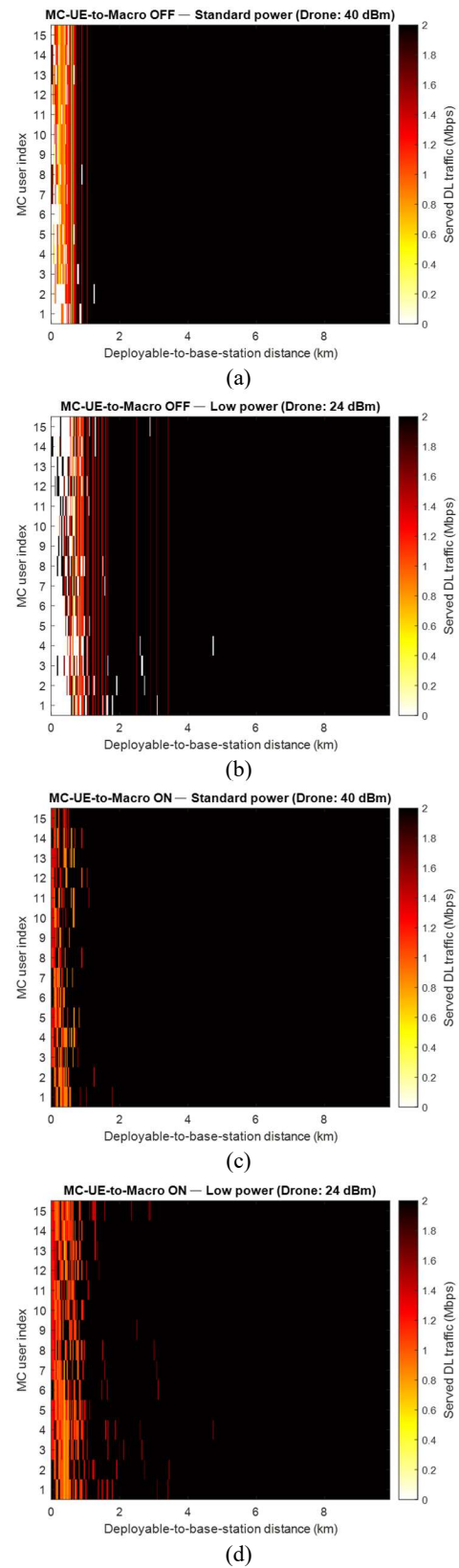

Fig. 6. Served DL traffic per MC user as a function of distance between deployable and macro base stations. In (a) and (b), MC users are not allowed to access the public network, whereas, in (c) and (d), they are. Also (a) and (c) consider deployable Tx power of 40 dBm, whereas, in (b) and (d), 24 dBm is used.

In this section, interference has been shown to have a major impact for short distances between public and deployable BSs. Hence, interference may be ignored in deployable network planning in, e.g., actual rural scenarios (forest, mountain, etc.) as was the case in the original scenario geometry, see Fig. 1. Besides, allowing MC users to access both the public and deployable infrastructures may be of great importance for ensuring optimal service at all times. Finally, note that similar interference impact and conclusions are to be extracted if the UAV deployable node was substituted with a truck deployable BS, i.e., scenario 3 in Section II. For the truck deployable BS, which employs a higher Tx power, more MC users will experience higher SINRs and, hence, they will connect to the public network/macro BS for shorter distances (between deployable and macro BSs) than in the UAV case. However, again, this will be detrimental to the normal users connected to the public network.

## IV. Conclusions and Future Work

In this article we analyze capacity and interference aspects in deploying a deployable BS to complement the existing public cellular network to provide on-demand temporary coverage for MC users in a rural scenario. Our simulation results indicate that the service performance of MC users depends on the type of deployable BS used and the configuration of the coexisting deployable and existing public cellular networks, i.e., macro BSs serve all the users or MC users only.

When deployable BSs are deployed far away from macro BSs belonging to the existing public cellular network, placing deployable BSs close to MC users can significantly improve the served traffic and SINR for MC users. It can also offload the MC users from the macro BSs, thereby freeing up resources from macro BSs to better serve normal users. In this case, the interference impact of adding deployable network to coexist with public cellular network is negligible due to the large distances between deployable and macro BSs.

When deployable BSs are located close to macro BSs, it can result in MC service disruption due to high interference levels between the deployable and public cellular networks. Hence, it is desirable to allow MC users to connect to the public network to benefit from the better link connection to the macro BS if compared to the link quality to the deployable BS. In this case, multi-network interference mitigation and prioritized MC traffic handling become very important aspects to ensure the performance requirements of MC users. This will be addressed in future work.